# Collective modes for an array of magnetic dots in the vortex state.

A. Yu. Galkin [1], B.A. Ivanov [2], and C.E. Zaspel [3]

[1] Institute of Metal Physics, National Academy of Sciences of Ukraine,

Vernadskii av. 36, 03142, Kiev, Ukraine

[2] Institute of Magnetism National Academy of Sciences and

Ministry of Education and Science of Ukraine, 36-B Vernadskii avenue,

03142 Kiev, Ukraine

[3] University of Montana-Western, Dillon, MT 59725, USA

**Abstract**

The dispersion relations for collective magnon modes for square-planar arrays of vortex-state magnetic dots, having closure magnetic flux are calculated. The array dots have no direct contact between each other, and the sole source of their interaction is the magnetic dipolar interaction. The magnon formalism using Bose operators along with translational symmetry of the lattice, with the knowledge of mode structure for the isolated dot, allows the diagonalization of the system Hamiltonian giving the dispersion relation. Arrays of vortex-state dots show a large variety of collective mode properties, such as positive or negative dispersion for different modes. For their description, not only dipolar interaction of effective magnetic dipoles, but non-dipolar terms common to higher multipole interaction in classical electrodynamics can be important. The dispersion relation is shown to be non-analytic as the value of the wavevector approaches zero for all dipolar active modes of the single dot. For vortex-state dots the interdot interaction is not weak, because, the dynamical part (in contrast to the static magnetization of the vortex state) dot does not contain the small parameter, the ratio of vortex core size to the dot radius. This interaction can lead to qualitative effects like the formation of modes of angular standing waves instead of modes with definite azimuthal number known for the insolated vortex state dot.

## 1. Introduction

During the last decade in the physics of magnetism much attention has been attracted to artificial materials, which are produced by means of modern nanotechnologies. Such materials often demonstrate unique physical properties, not found in bulk magnets. These properties attract much interest both from the viewpoint of fundamental physics and applications. Amongst these widely studied materials there are two-dimensional superstructures consisting of arrays of submicron magnetic particles (magnetic dots) on non-magnetic substrates. At the present time single particles can be produced with high accuracy, and also with a high level periodicity of particles in an array (see for recent review [1-3]). The magnetic dots are organized in square or rectangular structures, and are defined by rather high spatial regularity. From the point of view of dynamical properties this implies that for magnetic dot arrays the well-defined modes of collective oscillations characterized by a definite quasimomentum should exist.

Such systems possess a series of unique peculiarities. The distribution of magnetization in a single magnetic dot can be quite nontrivial, even for small dots. In the absence of an external magnetic field, a small enough non-ellipsoidal dot exhibits a single-domain nearly uniform magnetization state, the so called flower state or leaf state [6]. When the dot radius, $R$ is above a critical value, the vortex state with almost closed magnetic flux occurs [7]. This vortex state has been experimentally observed [8-18] for cylinder - shaped magnetic dots in the diameter range $2R = 100 \div 800$ nm and thickness range $L = 20 \div 60$ nm. Recent work [19] even reduced these values, demonstrating that the critical size can be as small as $L = 40$ nm and $R = 43$ nm, for permalloy. For such particles in a highly non-uniform state interesting peculiarities are expected related to the specific character of both eigenmodes for an isolated small particle and collective oscillations of an ordered system of dipolar coupled particles. The latter is caused by long distance character of interaction of magnetization oscillations localized on different particles. Models of magnetostatic moments with dipolar interaction have been theoretically studied for nearly 50 years, and many physical properties absent in exchange-coupled spin systems are known for these systems [20, 21]. This is also related to dynamics and it leads to peculiarities in spectra of oscillations of magnetic moments with dipolar coupling in a lattice [21, 22]. However, for dots in the vortex state the "rigid dipole" approach is not applicable and these results cannot be used directly.

Nevertheless, magnetic dot lattices exhibit physical properties, which are absent in traditional dipole-coupled systems. First, dot arrays in contrast to layered crystals are *literally* two-dimensional. Second, the characteristic energy of dipolar interaction of two magnetic dots can be comparable to or even higher than the thermal energy at room temperature [23, 24]. Next, and most important for the scope of this article, the isolated magnetic dot exhibits quite non-trivial discrete magnon mode spectra. The experiments performed in long coaxially magnetized ferromagnetic wires and in lattices of magnetic dots showed that the spin oscillations in such systems have discrete spectra that are a direct consequence of space quantization on a lateral surface of the dot (see for review [3]). During past few years, the essential progress was developed in the investigation of dynamical properties of vortex state dots made from soft magnetic materials. For them, quite nontrivial features, absent for fine particles with nearly-uniform magnetization, was found. For the following it is sufficient to mention the presence of a low frequency (sub-gigahertz) mode, describing small-amplitude precession of the vortex centre [25 – 33]. Also there is non-trivial dependence of the mode frequency on the principal and azimuthal wave numbers [34 - 42] as well the presence of doublet structure for some high frequency modes [27, 37, 30, 38]. Here the character of dot interaction is also nontrivial, and is a result of the fact that oscillations of the vortex state have almost closed magnetic flow, not described by the simple model of the "rigid" point dipole. These peculiarities naturally should manifest themselves in collective mode properties. Thus, magnetic dot arrays are specific materials with quite regular lattice structure, and long distance dipolar coupling between magnetic moments, which are rather large and manifest at high temperature.

In addition, fabrication and experimental study of magnetic dot arrays provide new physical systems for the testing of basic models of condensed matter physics. Basically, we can consider the magnetic dot lattice as an "artificial crystal", in which single magnetic dots serve as "atoms", or particles with complex discrete eigenspectra. The interaction of such "atoms" organized in a lattice, taking into account their identity and high regularity of the lattice, should result in the appearance of "energy bands", i.e. well defined collective modes. Obviously, here we have a classical wave problem, but with a formal analogy with the quantum mechanical problem. Each mode might arise from a single mode for a single dot; however, hybridization effects cannot be excluded. For such non-uniformly magnetized particles interesting peculiarities are expected related to the specific character of both eigenmodes for an isolated small particle and collective oscillations of an ordered lattice of dipolar coupled particles. The artificial character of these systems provides with a possibility to control independently the parameters of interaction and parameters similar to energies of atomic levels, by means of lattice and dot geometry modification.

This work is restricted to the case when the magnetic states are equivalent for all dots; in particular, the sense of rotation of the magnetization vector and vortex core polarization, are the same for various dots. The main goal of this article is the study of spectra of the collective oscillations for magnetic dot lattices, as a function of quasimomentum. Previously this problem has been considered numerically, without use of our main tool, the Bloch theorem, that automatically limits the authors working with the finite systems (large enough like $10^3$ dots [43] or even extremely small, with 9 dots only [44]).

The article is organized as follows. We will begin Section 2 with the formulation of a general approach valid for the description of dipolar interaction for the modes localized on different dots, with arbitrary distribution of magnetization, including dots in non-uniform states, arrayed in regular lattice. In this Section the brief description of the properties of the vortex state in the circular magnetic dot, as well as the classification of the normal modes for the isolated dot will be also presented. Then, the collective mode spectra originating from the two magnon modes for vortex state dots, vortex precessional mode (Section 3) and radially symmetrical mode, which can be called vortex breather mode (Section 4), will be calculated. Section 5 deals with the collective mode obtained as hybridization of a doublet known for higher non-radial modes. Section 6 summarizes the results obtained and gives the general conclusion.

**2. General static and dynamic properties of the vortex state circular dots.**

The magnetic dipole interaction for any distribution of magnetization, both for continuum ferromagnetic media or for patterned structure with ferromagnetic elements can be described by the following general equation [45, 46]

$$\hat{H}_{dipole} = \int \frac{\vec{M}(\vec{R}_1)\vec{M}(\vec{R}_2) - 3(\vec{M}(\vec{R}_1),\vec{R}_{12})(\vec{M}(\vec{R}_2),\vec{R}_{12})/|\vec{R}_{12}|^2}{2|\vec{R}_{12}|^3} d\vec{R}_1 d\vec{R}_2 , \qquad (1)$$

where the double integral over $\vec{R}$, which is the radius vector within all the volume of the system (see Fig. 1) appears, $\vec{R}_{12} = \vec{R}_1 - \vec{R}_2$, and $\vec{M}(\vec{R})$ is the magnetization. We can apply Eq.(1) to patterned media considering vector $\vec{M}$ as being non-zero only inside the ferromagnetic particles.

Let us now discuss general regularities typical for small oscillations of the magnetization of a fine magnetic particle with non-uniform ground state taking into account the condition $\vec{M}^2 = M_s^2$ is a constant inside the particle. The magnetization in the ground state, $\vec{M}_{(gr)}$ can be expressed as $\vec{M}_{(gr)} = M_s \vec{e}_3$, where $\vec{e}_3$ is a coordinate-dependent unit vector. This vector can be written through the angular variables for magnetization θ, φ which are introduced in the usual way, $\vec{e}_3 = \cos\theta\, \vec{e}_z + \sin\theta(\vec{e}_x \cos\varphi + \vec{e}_y \sin\varphi)$. Following [47], let us introduce a local Cartesian coordinate system $\vec{e}_3$, $\vec{e}_1$, $\vec{e}_2$, in which $\vec{e}_3$ plays a role of a quantization axis for the magnetization. For a non-uniform state the directions of the axis depend on coordinate $\vec{r}_l$ inside the given dot, see Fig. 1. Then, taking into account small deviations from $\vec{M}_{gr}$ the magnetization can be written as

$$\vec{M} = (M_S + \delta M_3)\vec{e}_3 + M_1 \vec{e}_1 + M_2 \vec{e}_2 . \qquad (2)$$

The projections of $M_1$ and $M_2$ define small oscillations of the magnetization, and the value $\delta M_3 \approx -(M_1^2 + M_2^2)/2M_S$ is quadratic over these small quantities. Oscillations of the magnetization can be considered in a framework of the classical Landau - Lifshitz equations. Within the framework of this classical approach not only the oscillations in a single dot but also the oscillations in a system of ordered dots can be described. Using the Landau-Lifshitz equation this problem is rather complicated when taking into account the general equation (1) for a patterned ferromagnet. As we will see, the analysis for the case of interest, the *regular in space* array consisting of *equivalent* dots for which the magnon eigenmodes are known, can be simplified considerably. The key points of the approach proposed here involve taking into account the translation symmetry of the array by use of the magnon formalism familiar in quantum theory of magnons in a crystal lattice, by the introduction of the Bose creation and annihilation operators, $a_\alpha^+$ and $a_\alpha$, for excitations of the α-th mode of oscillations of a given dot placed at the point $\vec{l}$ of the array,

$$M_1 = \sqrt{\mu_B M_s} \sum_\alpha [F_\alpha(\vec{r})a_\alpha^+ + F_\alpha^*(\vec{r})a_\alpha],$$
$$M_2 = i\sqrt{\mu_B M_s} \sum_\alpha [G_\alpha(\vec{r})a_\alpha^+ - G_\alpha^*(\vec{r})a_\alpha].$$
(3)

Here $\vec{r}$ is the radius vector in the coordinate system centered on the $\vec{l}$-th point as shown in Fig. 1, the functions $F_\alpha(\vec{r})$ and $G_\alpha(\vec{r})$ describes the magnetization distribution in the α-th mode of a single dot. In that case it is sufficient to take into account that in the classical micromagnetic problem the complex amplitudes of the given mode, $a_\alpha^* \propto e^{i\omega_\alpha t}$ and $a_\alpha \propto e^{-i\omega_\alpha t}$, where $\omega_\alpha$ is a mode frequency, correspond to the creation and annihilation operators, $a_\alpha^+$ and $a_\alpha$, respectively. Through these operators the Hamiltonian of non-interacting dots takes the trivial form

$$\hat{H}_0 = \sum_{\vec{l}} \hbar \omega_\alpha a_{\alpha,\vec{l}}^+ a_{\alpha,\vec{l}}.$$
(4)

Let us apply this approach to the vortex state magnetic dots. For thin enough dots in the vortex state the magnetization $\vec{M}$ can be considered to be independent of the *z*-coordinate along the normal to the dot, $\vec{M} = \vec{M}(r, \chi)$, where *r* and $\chi$ are the polar coordinates in the dot plane. Then the ground state magnetization inside the dot, which is in the $\vec{e}_3$ direction (see Eq. (2)) can be written as $\vec{M} = M_s \vec{e}_3$,

$$\vec{e}_3 = \vec{e}_z \cos\theta \pm \sin\theta \, (-\vec{e}_x \sin\chi + \vec{e}_y \cos\chi), \quad \theta = \theta(r), \qquad (5)$$

where the signs ± corresponds with two types of vortex states, with different sense of rotation of magnetization (vorticity). We will consider only the system with the same sign, say plus, for all the array. The function $\theta(r)$ differs from $\pi/2$ only in a small core region near the center of the dot $\Delta_0 < r \leq R$, where the characteristic length scale coincides with the value of the exchange length $\Delta_0$,

$$\Delta_0 = \sqrt{A/4\pi M_s^2}, \qquad (6)$$

$A$ is the inhomogeneous exchange constant. The value of $\Delta_0$ is approximately 5 nm for permalloy. Hence, in most of the dot the magnetization lies in the dot plane, which does not contribute to the total magnetization as it is compensated turning a $2\pi$ angle. The ground state magnetic moment is directed perpendicular to the plane of the dot and it is small compared with the saturated value of the dot. Again, the vortex core polarization can have two different signs, but we will consider the state with the same value of this quantity for all dots in the array.

For thick dots more general three-dimensional distribution should be considered. To the best of our understanding, it can be found only by direct numerical simulations, and we limit ourselves with much more simple two-dimensional distribution (5) with separation of coordinates. In polar coordinates with usual orts $\vec{e}_r$ and $\vec{e}_\chi$, $\vec{e}_r = \vec{e}_x \cos\chi + \vec{e}_y \sin\chi$, $\vec{e}_\chi = -\vec{e}_x \sin\chi + \vec{e}_y \cos\chi$, the ground state static distribution (5) takes the form $\vec{e}_3 = \vec{e}_z \cos\theta + \vec{e}_\chi \sin\theta$, and the unit vectors in (2) can be written as

$$\vec{e}_2 \equiv (\partial \vec{e}_3 / \partial \chi)/(\sin\theta) = -\vec{e}_r, \quad \vec{e}_1 \equiv \partial \vec{e}_3 / \partial \theta = -\vec{e}_z \sin\theta + \vec{e}_\chi \cos\theta, \qquad (7)$$

There has been much experimental and theoretical work regarding the measurement and calculation of the resonant frequencies of a single magnetic dot in the vortex state. An important point, common to vortex state dots and vortices in easy-plane two-dimensional magnets [48-50], is the normal mode classification. A full set of eigenfunctions for the ansatz (2) can be written as

$$M_1 = M_s \cdot f_{m,n}(r)\cos(m\chi + \omega_{m,n}t), \quad M_2 = M_s \cdot g_{m,n}(r)\sin(m\chi + \omega_{m,n}t), \tag{8}$$

where the azimuthal number, $m = 0, \pm1, \pm2$, has the sense of the principal number (knot number of $g_{m,n}$). For soft magnet vortex state dots, the characteristic frequency, which will frequently appear in the following, is $\omega_M = 4\pi\gamma M_s$, where $M_s$ is saturation magnetization, $\gamma = g\mu_B/\hbar$ is the gyromagnetic ratio, $g \approx 2$ is the Lande factor. For permalloy, the value of $\omega_M$ is approximately 30 GHz.

The form of the functions $f$ and $g$ depends on the numbers $n$ and $m$. The lowest mode with $n = 1$, $m = +1$ corresponds to the slow precession of the vortex core center around its equilibrium position (vortex precessional mode). This vortex mode has the form $g_m \propto -(1/r)\sin\theta(r)$ and $f_m \propto d\theta(r)/dr$ so that it is localized near the center of the system. Its frequency $\omega_{PM}$ is proportional to the dot aspect ratio $L/R$, with a small deviation from linear dependence as the aspect ratio, $L/R$ increases [26, 31-33]. For typical geometry for vortex state dots, $\omega_{PM}$ is in the sub GHz range and is much lower than the frequencies of all other modes, which also are smaller than $\omega_M$. This mode can be excited with a magnetic field pulse parallel to the dot plane. It was observed experimentally by many groups, by use of different experimental techniques, such as time resolved Kerr microscopy [28] and X-ray imaging [29, 33] and broadband ferromagnetic resonance [30].

The other modes for soft magnetic dots have higher frequencies of the order of a few GHz. The set of modes with $m = 0$ and different $n$ are associated with radially symmetric oscillations of the magnetization, owing to the fact that $M_{1,2}$ do not depend on $\chi$. These modes can be excited by a magnetic field pulse perpendicular to the dot plane. They also have been observed experimentally by use of Brillouin light scattering [37, 38] and time resolved Kerr microscopy with Fourier filtering [34-36]. Theoretical analysis [39-42] and numerical simulations were also used for investigations of these modes. For these modes, because of the absence of angular nodes, the contribution of the volume magnetic charges is maximal, thus their frequencies are usually higher than for non-radial modes with $|m|>1$ [35, 39, 42]. Also for these modes, approximate square root dependence of the frequency on the aspect ratio was found within both approximations, also coinciding with experiments [34-42].

The modes with higher values of numbers $m$ and $n$ were also observed and simulated numerically (see the nice figures of the mode profile in Ref. [38]). The common feature of all $m \neq 0$ modes, including higher modes with $m = \pm 1$, is the formation of doublets with $m = \pm |m|$, as a result of the scattering amplitude $\sigma_m(q)$. The splitting of the doublet is small compared to the average frequency of the doublet $(\omega_{|m|,n} - \omega_{-|m|,n}) \ll (\omega_{|m|,n} + \omega_{-|m|,n})$, and it contains the next power of the small parameter $L/R$ [40]. Note that this splitting can be strongly enlarged by application of the magnetic field perpendicular to the dot plane producing cone state vortices [47, 37]. The splitting is maximal for the higher modes with $m = \pm 1$, calculations show the linear dependence of $\Delta\omega_1$ on the aspect ratio $L/R$, as well as for the low frequency vortex precessional mode $\omega_{PM}$ (see Fig. 3 in Ref [40]). The higher modes with $|m|=1$ produce the high-frequency modulation of slow oscillations of the vortex coordinate as a function of time. For vortex state dots the beat is clearly seen in all the measurements of the vortex precessional motion [28-30], it was investigated in detail by use of Fourier analysis in Refs. [30, 38], and it was shown in [30] that two modes forming the doublet have a clear structure of angular *propagating* waves of the type (8) with opposite signs of *m*. In this nice experimental work, the doublet splitting was attributed to the vortex core: after making of a small (5 nm diameter) hole in the center of a sample, the doublet splitting disappear, and two angular *standing* waves of structure $g \propto \cos \chi \cos(\omega t)$ and $g \propto \sin \chi \cos(\omega t)$ forms instead of (8). On the other hand, the only angular *standing* waves were observed for larger dots with diameter 2-3 µm [35]. Also the same structure was obtained for $m = \pm 1$ modes by numerical simulations, with the frequencies difference about 0.2 GHz [38]. The much lower frequency difference can be attributed to steeper dependence of this parameter on dot diameter [39], which was also mentioned in [30]. Finally it is remarked that the doublet structure is very sensitive to the breaking of perfect circular symmetry, for example, cutting of the sample onto cuboids during numerical simulation procedure. We will discuss the transformation of the doublet of traveling waves to a pair of standing wave caused by arranging of circular dots to square lattice below in Section 5.

Thus, the mode structure for the insolated vortex sate magnetic dot is well established, mostly during the past four years. For most of them, the eigenfrequencies, as well as radial and angular dependencies are found by numerical simulations. As we mentioned above, the knowledge of these data, in fact, some simple integral characteristics of the magnetization distributions for given mode, allow one to describe the full spectrum of the collective modes for the dot array. Let us apply this general approach to the square lattice of vortex state dots, without a direct contact between dots, taking into account dipolar interaction of dots.

Now represent oscillation of the magnetization in one dot placed in the point $\vec{l}$ by means of Bose creation and annihilation operators $a^{+}_{\vec{l};m,n}$, $a_{\vec{l};m,n}$ for modes with given values *m* and *n* located on this dot. Using the correspondence (7) and the magnetization distribution (8) for a given mode and the general rule for substitution of time dependent exponents $\exp(i\omega_{m,n} t)$ or $\exp(-i\omega_{m,n} t)$ by the creation and annihilation operators $a^+_{m,n}$ or $a_{m,n}$ one obtains

$$\vec{M}_l(\vec{r},t) - M_s\vec{e}_3 = \sqrt{\mu_B M_s} \sum_{m,n} \{-ig_{m,n}(r)[e^{im\chi}a^+_{m,n} - e^{-im\chi}a_{m,n}](\vec{e}_x \cos\chi + \vec{e}_y \sin\chi) + \\ f_{m,n}(r)[e^{im\chi}a^+_{m,n} + e^{-im\chi}a_{m,n}][-\vec{e}_z \sin\theta(r) + (-\vec{e}_x \sin\chi + \vec{e}_y \cos\chi)\cos\theta(r)]\}, \quad (9)$$

where $r$, $\chi$ are polar coordinates $r_{\vec{l}}, \chi_{\vec{l}}$ taken in the coordinate system with the origin in $\vec{l}$-th dot center, $f_{m,n}(r)$, $g_{m,n}(r)$ are the functions for given mode, which we consider as known. Here the normalization of the functions $f_{m,n}(r)$ and $g_{m,n}(r)$ is chosen to satisfy the orthogonality conditions for vortex state modes $2\pi L \int f_{m,n}(r) g_{m',n'}(r) r dr = \delta_{n,n'} \delta_{m,m'}$ [49]. Then substitution $\vec{m}_{\vec{l}}$ and $\vec{m}_{\vec{l}'}$, $\vec{l} \neq \vec{l}'$, in the dipolar interaction formula (1) yields the interaction Hamiltonian for dots located at $\vec{l}$ and $\vec{l}'$ in the form of the sum of components describing the interaction of dots at the points $\vec{l}$ and $\vec{l}'$, $\hat{H} = \frac{1}{2}\sum_{l \neq l'} \hat{H}_{l,l'}$. The form of $\hat{H}_{l,l'}$ is simplified by taking into account only linear components in the Bose operators localized on different points, $a_{\alpha,l}$, $a^+_{\alpha,l}$ and $a_{\alpha,l'}$, $a^+_{\alpha,l'}$, here $\alpha = \{n,m\}$. For example, one of those components is expressed as follows.

$$\hat{H}_{1,2} = \int \frac{\vec{M}_{\vec{l}_1}\vec{M}_{\vec{l}_2} - 3(\vec{M}_{\vec{l}_1}\vec{L}_{12})(\vec{M}_{\vec{l}_2}\vec{L}_{12})/|\vec{L}_{12}|^2}{|\vec{L}_{12}|^3} d\vec{r}_1 d\vec{r}_2, \quad (10)$$

where $\vec{L}_{12} = \vec{l}_1 - \vec{l}_2 + \vec{r}_1 - \vec{r}_2$, $\vec{M}_{\vec{l}}$ is the magnetization of the form (9) linear in $a_{\alpha,l}$, $a^+_{\alpha,l}$ for the $\vec{l}$-th magnetic dot. The Hamiltonian $\hat{H}_{1,2}$ splits into the sum of components describing the interaction $\alpha$-th mode localized on the magnetic dot $\vec{l}_1$, and $\beta$-th mode localized on $\vec{l}_2$. The calculation of the integrals entering this Hamiltonian is simplified by use of inequalities $R \ll a$ and $\Delta_0 \ll R$. In light of this we will expand the interaction (10) in the small parameter $R/|\vec{l} - \vec{l}'|$, which is just the magnetic multipole expansion. Finally, we will arrive at the expression of interaction describing the mode with numbers $\alpha = \{n,m\}$ localized on dot $\vec{l}$, and the mode with numbers $\beta = \{n',m'\}$ localized on dot $\vec{l}'$ in the form

$$A^{(\alpha,\beta)}(\vec{l}-\vec{l}')a_\alpha a_\beta^+ + B^{(\alpha,\beta)}(\vec{l}-\vec{l}')a_\alpha^+ a_\beta^+ + h.c. \qquad (11)$$

Here an essential point is that the coefficients $A^{(\alpha,\beta)}$ and $B^{(\alpha,\beta)}$ depend only on the radius-vector $\vec{l}_1 - \vec{l}_2$ connecting the centers of the dots 1 and 2. Also for the dipolar approximation, $A$, $B$ $\propto 1/|\vec{l}-\vec{l}'|^3$.

Next a direct application of the Bloch theorem allows the collective modes to be introduced via states $a_k^+$, $a_k$,

$$a_l = \frac{1}{\sqrt{N}}\sum_{\vec{k}} a_k e^{i\vec{k}\vec{l}} \quad \text{and} \quad a_l^+ = \frac{1}{\sqrt{N}}\sum_{\vec{k}} a_k^+ e^{-i\vec{k}\vec{l}}, \qquad (12)$$

which are characterized by a definite quasi-momentum $\vec{k}$, defined to within the reciprocal lattice vector, $\vec{g}$. For the square array $\vec{l} = a(\vec{e}_x l_x + \vec{e}_y l_y)$ and $\vec{g} = 2\pi(\vec{e}_x n_x + \vec{e}_y n_y)/a$, where $l_x, l_y$ and $n_x$, $n_y$ are integers. After use of the equality $\sum_l e^{i\vec{k}\vec{l}} = N\delta(\vec{k})$, the Hamiltonian of the interacting modes reads

$$\hat{H} = \frac{1}{2}\sum_{\vec{k}}\sum_{\alpha,\beta}[A_k^{(\alpha,\beta)} \cdot a_{\alpha,k}^+ a_{\beta,k} + B_k^{(\alpha,\beta)} \cdot a_{\alpha,k}^+ a_{\beta,-k}^+ + h.c.], \qquad (13)$$

where the coefficients are expressed via the lattice sums,

$$A_k^{(\alpha,\beta)} = \sum_{\vec{l}\neq 0} A^{(\alpha,\beta)}(\vec{l}) \cdot e^{i\vec{k}\vec{l}}, \quad B_k^{(\alpha,\beta)} = \sum_{\vec{l}\neq 0} B^{(\alpha,\beta)}(\vec{l}) \cdot e^{i\vec{k}\vec{l}}. \qquad (14)$$

Note that only for modes with $m = 0$ or $m = \pm 1$, do the integrals $\int \delta \vec{M}(\vec{r},t) d\vec{r}$ differ from zero, or only these modes correspond to oscillations of the total magnetic moment, $\delta \vec{m}(t)$ of an isolated magnetic dot. In this case the Hamiltonian for α-th collective mode takes the following form:

$$\hat{H} = \sum_{\vec{k}} [A_k^\alpha \cdot a_{\alpha,k}^+ a_{\alpha,k} + \frac{1}{2}(B_k^\alpha \cdot a_{\alpha,k}^+ a_{\alpha,-k}^+ + h.c.)], \quad (15)$$

where $A_k^\alpha \equiv A_k^{(\alpha,\alpha)}$ and $B_k^\alpha \equiv B_k^{(\alpha,\alpha)}$ are the "diagonal" over α parts of general Hamiltonian (13). It can be diagonalized by the usual Bogolyubov $u$-$v$ transform, (see for example [45]) and the dispersion law reads

$$\hbar \omega_\alpha (\vec{k}) = \sqrt{(A_k^\alpha)^2 - (B_k^\alpha)^2} \quad , \quad (16)$$

where $\hbar$ is a Planck's constant.

For the $m = 0$ mode the moment is directed perpendicular to the dot plane, whereas for the $m = \pm 1$ modes it lies in the dot plane. Hence, the coupling coefficient of modes with $m = 0$ and $m = \pm 1$, in the lowest approximation in $1/|\vec{l} - \vec{l'}|$ becomes zero. The dipole interaction is the main source of the dispersion for collective modes for $|m| = 1$, which is considered in detail in Section 3 and Section 5. On the other hand, for adequate description of mode collectivization some terms of non-dipolar interaction are important. As we will show below in Section 4, non-dipolar terms gives an essential contribution to the dispersion for modes with $m = 0$ having non-zero magnetic moment. The coupling can also be present for modes with non-equal values of $m$. This interaction is especially important for modes with $m = \pm |m|$ which form doublets. For the collective modes originating from such doublets, the diagonalization of the Hamiltonian (13) has to be done taking into account two modes of different $m$. We will discuss this below for an important example, namely higher $m = \pm 1$ modes having a strong interaction of the dipole type (see Section 5). It is remarked that some coupling coefficients of modes with $|m| \neq |m'|$ appear to be nonzero. For instance, this is the case for modes with $m = 0$ and $m = \pm 2$ with non-dipolar interaction proportional to $1/|\vec{l} - \vec{l'}|^5$. The presence of such a coupling allows the indirect excitation of modes with $|m| > 1$ through direct excitation of modes with $m = 0, \pm 1$ under influence of a uniform alternating magnetic field.

### 3. The mode of collective precession of vortices

The mode of precessional oscillations of the vortex in a single dot has the lowest frequency, $\omega_{PM}$. For this mode, the functions $f(r)$ is localized near the vortex core, see [26, 27],

$$f_{PM}(r) = B \cdot d\theta/dr, \tag{17}$$

and the calculation of the value of normalizing coefficient from the orthogonality condition does not depend on the details of the function $g_{PM}(r)$, giving $B = 1/\sqrt{2\pi L}$. Neglecting the small terms arising from local functions like $\cos\theta$ and $d\theta/dr$, the oscillating part of the magnetization inside the dot can be written through the function $g(r)$ only, $M_x = -M_2 \cos\chi$, $M_y = -M_2 \sin\chi$, and $M_2 = i\sqrt{\mu_B M_s} [a^+ \cdot \exp(i\chi) - a \cdot \exp(-i\chi)]$. Then, non-zero components of the total magnetic moment are

$$m_x = -i\pi[a^+ - a] L \sqrt{\mu_B M_s} \int_0^R rg(r)dr, \quad m_y = \pi[a^+ + a] L \sqrt{\mu_B M_s} \int_0^R rg(r)dr, \tag{18}$$

and the interaction Hamiltonian can be expressed as

$$H_{PM}^{(int)} = \frac{2\mu_B \pi L R^2 M_s}{9a^3} \sum_k \left[ -\sigma(\vec{k}) a_k^+ a_k + \frac{3}{2} \left( \sigma_c(\vec{k}) a_k^+ a_{-k}^+ + \sigma_c^*(\vec{k}) a_k a_{-k} \right) \right], \tag{19}$$

where $\sigma(\vec{k})$ and $\sigma_c(\vec{k})$ are two dipole sums,

$$\sigma(\vec{k}) = \sum_{\vec{l} \neq 0} \frac{1}{(l_x^2 + l_y^2)^{3/2}} e^{i\vec{k}\vec{l}}, \quad \sigma_c(\vec{k}) \equiv \sigma'(\vec{k}) + i\sigma''(\vec{k}) = \sum_{\vec{l} \neq 0} \frac{(l_x - il_y)^2}{(l_x^2 + l_y^2)^{5/2}} \cdot e^{i\vec{k}\vec{l}}. \qquad (20)$$

The same sums will appear below in Section 5, for the description of higher |m|=1 modes, and they are also important for the collective modes for dots with homogeneous magnetization [22]. The total Hamiltonian takes the form $\hat{H}_{PM} = \sum_k \omega_{PM} a_k^+ a_k + \hat{H}_{PM}^{(int)}$, where $\omega_{PM}$ is the frequency of the vortex precessional mode for the isolated dot. The sum $\sigma(\vec{k})$ at $\vec{k} = 0$ has the finite value $\sigma(0) = 9.03362$ while $\sigma_c(0)$ is equal to zero, as seen in Fig. 2. It is important to note their non-analytical $\vec{k}$ dependence at small values of $\vec{k}$,

$$\sigma(\vec{k}) = \sigma(0) - kF(\vec{k}),$$

$$\sigma_c(\vec{k}) \equiv \sigma'(\vec{k}) + i\sigma''(\vec{k}) = \frac{(k_x - ik_y)^2}{|\vec{k}|} \cdot G(\vec{k}), \qquad (21)$$

where $k = |\vec{k}|$ and the functions $F(\vec{k})$ and $G(\vec{k})$ are analytic as $k \to 0$, with the limit values $F(\vec{k}) \to 2\pi a$ and $G(\vec{k}) \to 2\pi a/3$ as $\vec{k} \to 0$. Both functions, $F(\vec{k})$ and $G(\vec{k})$, are invariant with respect to the symmetry group of the dot lattice.

Note also that the ratio of frequencies $\omega_0/\omega_M$, for the isolated dot with $\omega_0 \equiv \omega_{PM}$, which is one of small parameters of the problem for the $m = 0$ mode (see next Section) is not present in the interaction Hamiltonian. This feature is very important because the value of frequency $\omega_0$ for the isolated dot is especially small for the precessional mode. The interaction appears as only one small parameter $\pi L R^2/a^3$ defined by the geometry of the system.

The Hamiltonian $\hat{H}_{PM}$ can be diagonalized by use of general equation (16) and the frequency of the collective precessional motion of the vortices located on different dots in the array is

$$\omega_{PM}(\vec{k}) = \sqrt{[\omega_0 + \omega_{int}\Sigma_{(+)}(\vec{k})][\omega_0 + \omega_{int}\Sigma_{(-)}(\vec{k})]}, \qquad (22)$$

where

$$\Sigma_{(\pm)}(\vec{k}) = \sigma(0) - \sigma(\vec{k}) \pm 3|\sigma_c(\vec{k})|, \qquad (23)$$

$\omega_0 = \omega_{PM} - \omega_{int}\sigma(0)$, the value of $\omega_0$ has the sense of the gap for this mode, $\omega_{int} = \omega_M L R^2/36a^3$, and as before $\omega_M = 4\pi\gamma M_s$. It is worth noting here, the value of $\omega_0$ could be negative for large enough $\omega_{int}$, that describes the instability of the vortex state array with respect to the transition to in-plane.

The character of this dispersion is determined by the ratio of two parameters, $\omega_{PM}$ describing the frequency of vortex core slow precession for isolated dot, and $\omega_{int}$, determined by the lattice geometry. For the value of $\omega_{PM}$, we can use the theoretical estimate in the form $\omega_{PM} \approx C \cdot \omega_M(L/R)$ [26, 31,33], which agrees well with the experiments [28 – 30,33] with the value of the numerical constant $C \sim 0.15$. Thus, both characteristic frequencies $\omega_{PM}$ and $\omega_{int}$ are directly proportional to the dot thickness $L$, which is present in the dispersion law as a common multiplier, and the ratio of all terms is determined by only one geometrical parameter, which is just an array density given by $R/a$. Then, simple analysis shows, that for reasonable values of $2R/a < 1$ the frequency of collective mode $\omega(\vec{k})$ is growing monotonically with $|\vec{k}|$ inside all of the Brillouin zone (see Figure 3). The predicted value of dispersion can be non-small, of order 20% for the dense array.

Note that the magnon spectra have a peculiarity at small $\vec{k}$,

$$\omega_{PM}(\vec{k}) = \omega_0\left(1+|\vec{k}|R^3/36Ca^2\right) = \omega_0 + 2\pi\gamma M_s |\vec{k}| L (\pi R^2/9a^2) \text{ as } \vec{k} \to 0, \qquad (24)$$

where the value of $\omega_0$ has the sense of the gap for this mode.

### 4. Collective modes for oscillations with $m = 0$.

For the $m = 0$ mode the oscillating part of the total magnetic moment for any dot is parallel to the $z$-axis, with $\delta\vec{m} = \vec{e}_z \delta m_z$. It can be written using the function $f_0(kr)$ only,

$$\delta m_z = 2\pi L \int_0^R M_1 \sin\theta r dr \left(a_0^+ + a_0\right) \cong 2\pi A_0 I_0 L R^2 \sqrt{\frac{\omega_0}{\omega_M}}\sqrt{\mu_B M_s}\left(a_0^+ + a_0\right), \qquad (25)$$

where $A_0$ is the value of normalizing coefficient for $m = 0$, $A_0^{-2} = \pi L R^2 |J_1(j_{n,0})J_1(j_{n,0})|$, $\omega_0$ is the frequency for the isolated dot, $I_0 = \int_0^1 J_0(j_{n,0}\rho)\rho d\rho = J_1(j_{n,0})/j_{n,0}$, $a_0^+, a_0$ are operators of creation and annihilation for the $m = 0$ mode, and $j_{n,0}$ is the root of the Bessel function, $J_0(z)$. For the principal $n = 1$ mode it is a smallest root. Then, after using the Bloch theorem the interaction Hamiltonian for the $m = 0$ mode within the dipolar approximation can be written in the form

$$H_{m=0}^{(\text{dipole})} = \frac{4\mu_B \pi L R^2 M_s}{j_0^2 a^3}\frac{\omega_0}{\omega_M}\sum_k \sigma(\vec{k})\left[a_k^+ a_k + \frac{1}{2}\left(a_k^+ a_{-k}^+ + a_k a_{-k}\right)\right], \qquad (26)$$

where $\sigma(\vec{k})$ is the dipole sum introduced in (20). The coupling coefficient here is proportional to the dynamical part of the total $z$-projection of the dot magnetization. Notice that in contrast to the static magnetization of the vortex state dot, the dynamical part does not contain the small parameter, $\Delta_0/R$. On the other hand, the important feature of the collective $m = 0$ modes is the proportionality of the dipole interaction to the frequency, $\omega_0$ of the isolated dot. Due to the presence of the small parameter $\omega_0/\omega_M$, the non-dipole interaction proportional to the function $g(r)$ can be also important here. The $m = 0$ mode corresponds to "breathing" of the magnetization, with the oscillations of the vortex core size without change of the angular dependence (5). Thus, the dynamical part of magnetization leading to zero in-plane magnetic moment and non-dipolar interaction can be written through the function $g(r)$, $\delta \vec{M}(r,t) = i\sqrt{\mu_B M_S} \cdot g(r)(\vec{r}/r)(a^+ - a)$. Inserting this function into the general Hamiltonian (10) we can find that in the first (dipolar) approximation this interaction vanishes due to integration over polar coordinates, $\chi_1$ and $\chi_2$. But after expansion over $\vec{r}_1/a$ and $\vec{r}_2/a$, the non-zero contribution appears from the next powers of these quantities. Doing this expansion and calculation of the integrals over $d\chi_1$ amd $d\chi_2$, we can find the non-dipole contribution to the interaction Hamiltonian of the form

$$\hat{H}_{m=0}^{(\text{non-dipole})} = 2\mu_B M_S (4\pi\eta) \cdot \left(\frac{R^4 L}{a^5}\right)\left(\frac{\omega_M}{\omega_0}\right) \sum_{\vec{k}} \sigma_2(\vec{k}) [a_k^+ a_k - \frac{1}{2}(a_k^+ a_{-k}^+ + a_k a_{-k})]. \quad (27)$$

Here note that the presence of higher (in comparison with the purely dipole contribution $\hat{H}_{m=0}^{(dip)}$) powers of $R/a$, which is small for the low density array, as well as the presence of the large parameter $(\omega_M/\omega_0)$, in contrast with small parameter $(\omega_0/\omega_M)$ in the dipolar approximation (26). In this equation the sum $\sigma_2(\vec{k})$,

$$\sigma_2(\vec{k}) = \sum_{|\vec{l}|\neq 0} \frac{e^{ik\vec{l}}}{(l_x^2 + l_y^2)^{5/2}}, \quad (28)$$

appears, which is one of the higher dipole sums $\sigma_{2p}$ discussed in the Appendix. Here $\eta$ is a numerical coefficient of order unity. For the fundamental $m = 0$ mode $\eta \cong 1.20514$; for the higher modes with $m = 0$ its value is even smaller.

The total interaction Hamiltonian is the sum of two terms, $\hat{H}_{m=0}^{(dipole)}$ and $\hat{H}_{m=0}^{(non-dipole)}$. Both have the structure of the general Eq. (15) with the simple connection of A and B, $A_k^{(d)} = B_k^{(d)}$ for the dipole part and $B^{(nd)} = -A^{(nd)}$ for non-dipole part. Following the general Eq. (16) we can write $\omega^2(\vec{k}) = (\omega_0 + A^{(d)} + A^{(nd)})^2 - (A^{(d)} - A^{(nd)})^2$. Finally, the frequency of collective modes takes the form of the product of two terms, the first having the dipolar origin, and the second a higher multipole origin,

$$\omega(\vec{k}) = \sqrt{[1 + \frac{LR^2}{a^3 j_0^2}\sigma(\vec{k})][\omega_0^2 + 2\omega_M^2 \eta \cdot \frac{LR^4}{a^5}\sigma_2(\vec{k})]} \; . \tag{29}$$

In the long wave (small $|\vec{k}|$) regime the main origin of dispersion is determined by non-analytical behavior governed by the first bracket of dipole origin, and in this limit case

$$\omega = \omega_{gap}\left(1 - \frac{\pi LR^2}{a^2 j_0^2}|\vec{k}|\right) \; , \quad \omega_{gap}^2 = \omega_0^2 + 2\omega_M^2 \eta \cdot \frac{LR^4}{a^5}\sigma_2(0) \; , \tag{30}$$

where $\omega_{gap}$ is a frequency at $|\vec{k}| = 0$. It is obvious that the linear term is the most important for small $|\vec{k}|$, and other terms produce only quadratic over $|\vec{k}|$ analytical behavior. Here the quantity $\omega_{gap}$ becomes a common multiplier, and the dispersion is described by the purely geometrical parameter, $\pi LR^2/a^3$ which is simply the ratio of the dot volume and cube of the lattice size. Notice that the dispersion here is not small, and it does not contain the small parameters $\Delta_0/R$ or $\omega_0/\omega_M$.

Using the numerical data for the sums $\sigma(\vec{k})$ and $\sigma_2(\vec{k})$ it is easy to describe the dispersion relation for different directions of wave vector $\vec{k}$. The dispersion law for two definite directions of the wave vector, $\vec{k} \parallel (1,1)$ and $\vec{k} \parallel (1,0)$ is illustrated in the Fig. 4. For a thin enough dot, the dispersion relation can be simplified by use of a square root dependence of $\omega_0$, $\omega_0/\omega_M \approx C \cdot \sqrt{L/R}$ [36, 37, 41], with the constant $C \approx 1.1$, which is in good agreement with experiments. Using this formula, the dispersion law can be rewritten through purely geometrical factors with the frequency for isolated dot, $\omega_0$ as a common multiplier,

$$\omega(\vec{k}) = \omega_0 \sqrt{[1+\frac{LR^2}{a^3 j_0^2}\sigma(\vec{k})][1+2\eta \cdot \frac{R^5}{a^5 C^2}\sigma_2(\vec{k})]} \tag{31}$$

Thus, the common behavior of the dispersion law is determined by the geometry of the dot array. Since both functions, $\sigma(\vec{k})$ and $\sigma_2(\vec{k})$ are decreasing functions of $|\vec{k}|$ with the minimal value at the point of the type (1,1) with $\vec{k} = \vec{k}_0 = \pi(\vec{e}_x + \vec{e}_y)/a$, the minimal value of $\omega^2(\vec{k})$ is also present at these points. One can see that for thin dots with $L \ll R$ the dispersion connected to the first bracket in Eq. (31) is small. For the second bracket in Eq. (31) the small dot aspect ratio $L/R$ is not present, but other geometrical limitations appear. In particular, the dot diameter $2R$ has to be smaller than the array spacing $a$. Taking the value of $\eta \approx 1.1$ and the minimal value of $\sigma_2(\vec{k}_0) \approx \sqrt{2} - 4 \approx -3.29$ (see Appendix) of $m = 0$ modes for vortex state dots at $2R/a \geq 1$ the dispersion term in the second brackets is approximately $1/6$ smaller than unity. Thus, for vortex state dots the dispersion for the collective mode with $m = 0$ is usually small. For example, for the typical dot geometry [19-21], where the thickness $L = 100$ nm with the dot diameter $2R = 1$ μm even for the dense array of spacing $a \sim 2R$, the first bracket is practically unity. For this case even the non-analytic behavior of the type $\omega_g - \omega \propto |\vec{k}|$ predicted above is not clearly seen in the scale used on the main part of Fig. 4. Surely, the non-analytical dependence of $\omega(\vec{k})$ has to be important at small enough $|\vec{k}|$, and it is well seen with a change of scale (see insert on this figure). On the other hand, the vortex state can be present for thick magnetic dots with $L \sim 2R$, and even for the particles elongated along the dot axis. We believe our theory can be valid, at least qualitatively, for this case. With increasing of the aspect ratio the frequency of isolated dot become *smaller* than predicted by the simple formula $\omega_0/\omega_M \approx C\sqrt{L/R}$. For large enough aspect ratio, $L/R$ the dispersion caused by the first bracket can be stronger.

**5. The interaction between modes with different *m***

The dot lattice symmetry is low compared to a single circular dot; therefore, modes with different asimuthal numbers *m* localized on different dots can interact with each other. Analyzing the formation of collective modes this interaction in principle should be taken into account. One can expect that the interaction effect is most prominent for those oscillations for which the frequencies of local modes are close in the case of the single dot. For the vortex state, the latter are doublets created by the modes with $m = \pm|m|$. The splitting of these doublets is connected to the gyroscopic character of the magnetization dynamics, which is usually small, and it is the most pronounced for the $m=\pm 1$ modes [39].

The regularities of interaction between modes with different *m* will be demonstrated by the example of $|m|=1$ modes. This case is interesting because it exhibits the extremely low frequency vortex precessional mode, for these modes the magnetic dipole interaction is maximal, and finally the higher $|m|=1$ mode was recently observed on the slow vortex precessional motion by time-resolved Kerr microscopy [30, 38].

Let us consider the interaction of modes with $m = +1$ and $m = -1$ localized on different dots of the array, and we will discuss the formation of two branches of collective excitations from those modes. For a given dot the alternating part of the magnetization $\delta \vec{M}$ can be represented through the annihilation and creation operators for two types of modes,

$$\delta \vec{M}(r,t) = -i\sqrt{\mu_B M_s}(\vec{e}_x \cos \chi + \vec{e}_y \sin \chi) \times$$
$$[g_{m=1}(r)(e^{i\chi}a^+ - e^{-i\chi}a) + g_{m=-1}(r)(e^{-i\chi}\bar{a}^+ - e^{i\chi}\bar{a})] \quad . \tag{32}$$

Here to reduce the length of the expression we omitted the local (proportional to cosθ, which is non-zero near the vortex core only) part of $\delta \vec{M}$ and denote $a \equiv a_{m=1}$, $\bar{a} \equiv a_{m=-1}$, and the same symbols will be used below for the Fourier components of the operators $a$, $a^+$.

After simple transformations and application of the Bloch theorem the Hamiltonian of interaction reads

$$H^{(int)}_{|m|=1} = \sum_k \sigma(\vec{k})[-G^2 a_k^+ a_k - \bar{G}^2 \bar{a}_k^+ \bar{a}_k + G\bar{G}(a_k^+ \bar{a}_{-k}^+ + a_k \bar{a}_{-k})] -$$
$$- \frac{3}{2}\sum_k \left[ \sigma_c(\vec{k}) \cdot (G^2 a_k^+ a_{-k}^+ + \bar{G}^2 \bar{a}_k^+ \bar{a}_{-k}^+ - 2G\bar{G}a_k^+ \bar{a}_k) + h.c. \right], \tag{33}$$

note presence of the dipole sums (20), where we also introduced the notations

$$G = \pi L\sqrt{\mu_B M_s/a^3}\int_0^R rdr g_{m=1}(r) \;,\; \bar{G} = \pi L\sqrt{\mu_B M_s/a^3}\int_0^R rdr g_{m=-1}(r) \;. \tag{34}$$

Thus in order to analyze this case the Hamiltonian can be written via the previous dipole sums, σ($\vec{k}$) and σ_c($\vec{k}$). Diagonalization of the total Hamiltonian can be carried out by means of a simple generalization of the *u–v* Bogolyubov transformation onto the case of two interacting modes. Denoting the frequencies of modes with $m = 1$ and $m = -1$ for the single dot as $\omega_0$ and $\bar{\omega}_0$, respectively, after long but simple algebra the dispersion relation for two branches of collective oscillations can be written as

$$\left(\omega_k^2 - \Omega^2\right)\left(\omega_k^2 - \bar{\Omega}^2\right) + 2\omega_k^2 G^2 \bar{G}^2 \left(\sigma^2 - |3\sigma_c|^2\right) - G^2 \bar{G}^2 \left\{2\omega_0 \bar{\omega}_0 \left(\sigma^2 + |3\sigma_c|^2\right) - 2\sigma\left(\bar{\omega}_0 G^2 + \omega_0 \bar{G}^2\right)\left(\sigma^2 - |3\sigma_c|^2\right) + G^2 \bar{G}^2 \left(\sigma^2 - |3\sigma_c|^2\right)^2\right\} = 0$$
(35)

where $\Omega^2 = \left(\omega_0 - G^2 \sigma\right)^2 - G^4 |3\sigma_c|^2$, $\bar{\Omega}$ denotes the replacement of $G \to \bar{G}$ and $\omega_0 \to \bar{\omega}_0$, and to shorten the formula the notations $\sigma = \sigma(\vec{k})$, $\sigma_c = \sigma_c(\vec{k})$ are used.

The structure of Eq. (35) is clear: The two brackets in the first term reproduces the dispersion laws of collective modes $\omega_1(\vec{k}) = \Omega$ and $\omega_2(\vec{k}) = \bar{\Omega}$, which in turn might be obtained by taking into account the modes of the same type localized on different dots. The formulae for $\Omega$ and $\bar{\Omega}$ have the same structure as Eq. (22) for the vortex precessional mode. The use of simple expressions $\omega_1(\vec{k}) = \Omega$ and $\omega_2(\vec{k}) = \bar{\Omega}$ corresponds physically to "disconnection" or noninteraction of one of the initial modes having $m = |m|$ or $m = -|m|$ from the interaction with the other mode. Formally such disconnection can be obtained by considering the limits $G/\bar{G} \to 0$ or $\bar{G}/G \to 0$, when in Eq. (35) only the first item remains. The same result appears also for the large difference of $\omega_0$ and $\bar{\omega}_0$.

The form of the dispersion relation in Eq. (35) is a simple biquadratic expression and its explicit solution can be obtained easily, but this solution appears quite cumbersome. Therefore, the results of the analysis become more clear if one uses inequalities corresponding to the cases we are interested in. Let us consider at the beginning collectivization of modes incorporated in the doublet. This corresponds to the condition $\omega_1(\vec{k}) - \omega_2(\vec{k}) \ll \omega_{1,2}(\vec{k})$, and to approximate equality of the characteristic dipole moments, $G$ and $\bar{G}$. The frequencies $\omega_1(\vec{k})$ and $\omega_2(\vec{k})$ are not small for the doublet while the interaction can be considered small. Then in this approximation on the small parameters $\omega_1(\vec{k}) - \omega_2(\vec{k})$ and $G$ the following result is obtained

$$\omega_{1,2}^2(\vec{k}) = \frac{1}{2}\left(\Omega^2 + \overline{\Omega}^2\right) - G^2\overline{G}^2\left(\sigma^2 - |3\sigma_c|^2\right) \pm$$
$$\pm \frac{1}{2}\sqrt{\left(\Omega^2 - \overline{\Omega}^2\right)^2 + 4G^2\overline{G}^2(\omega_0 + \overline{\omega}_0)^2 |3\sigma_c|^2} \quad . \tag{36}$$

Even after this simplification the character of the dispersion law of modes is not clear. To make it transparent, we have to use one more inequality considering different connections between small parameters $\Omega - \overline{\Omega} \approx \omega_0 - \overline{\omega}_0$ and $G^2$. For $\omega_0 - \overline{\omega}_0 > \sigma G^2$ one can easily obtain

$$\omega_1^2(\vec{k}) = \Omega^2 - \sigma^2 G^2 \overline{G}^2 + 2\omega_0 |3\sigma_c|^2 G^2 \overline{G}^2 / (\omega_0 - \overline{\omega}_0),$$
$$\omega_2^2(\vec{k}) = \overline{\Omega}^2 - \sigma^2 G^2 \overline{G}^2 - 2\overline{\omega}_0 |3\sigma_c|^2 G^2 \overline{G}^2 / (\omega_0 - \overline{\omega}_0). \tag{37}$$

In this expression, the deviation of the dispersion law from simple expressions $\omega_1(\vec{k}) = \Omega$ or $\omega_2(\vec{k}) = \overline{\Omega}$, which corresponds the formation of collective modes from the one local mode only, is small. It contains an additional degree of the small parameter, $\sigma G^2/(\omega_0 - \overline{\omega}_0)$ in comparison with contributions from interaction of modes with the same $m$, which are already present in the equations $\omega_{1,2}(\vec{k}) = \Omega$ or $\overline{\Omega}$. Thus, if the initial doublet splitting is larger than the mode interaction, in the first approximation on the mode interaction the collectivization of modes with different $m$ can be considered separately, by use of simple expressions

$$\omega_1(\vec{k}) = \omega_0 - \sigma(\vec{k})G^2 \text{ or } \omega_2(\vec{k}) = \overline{\omega}_0 - \sigma(\vec{k})\overline{G}^2 \text{ for } \omega_0 - \overline{\omega}_0 > \sigma G^2. \tag{38}$$

In fact, the omitted terms in $\Omega$ or $\overline{\Omega}$ here are of the same order of magnitude as the items describing the sign-dependent corrections in (37) and have to be omitted.

For other the limit case, if the difference of the frequencies is extremely small, $\omega_0 - \overline{\omega}_0 \ll \sigma G^2$ the splitting is still non-zero even in the limit case $\omega_0 = \overline{\omega}_0$ and $G = \overline{G}$. For this last limit the dispersion laws for two modes can be presented as

$$\omega_{1,2}^2(\vec{k}) = \omega_0^2 - 2\omega_0 G^2 [\sigma(\vec{k}) \pm |3\sigma(\vec{k})|] \text{ for } \omega_0 - \overline{\omega}_0 \ll \sigma G^2. \tag{39}$$

These dispersion laws have nothing to do with $\Omega(\vec{k})$ or $\bar{\Omega}(\vec{k})$. This is clearly seen after the comparison of the expression (39) with the approximate formula (38) written with the same accuracy as (38). The equation (39), written for collective modes with small difference of $\omega_0 - \bar{\omega}_0 \ll \sigma G^2$, contains the complex sum $\sigma_c(\vec{k})$, describing the direct coupling of modes with $m = 1$ and $m = -1$ of the type $a_k^+ \bar{a}_k$ (see Eq. (33)). The analysis shows, for this case, the normal mode found by generalized $u$–$v$ Bogolyubov transformation contains a combination of the operators $a_k$ and $\bar{a}_k$ with approximately equal amplitudes. Thus, for the case of extremely small $\omega_0 - \bar{\omega}_0$, the distribution of magnetization for the collective modes have principally different angular dependence than for non-interacting modes in the doublet. The collective modes are *not* the modes with well-defined value of the azimuthal number $m$ (angular propogating waves) discussed above. Each of two modes originating from the doublet with $m = \pm |m|$ at $\omega_0 - \bar{\omega}_0 \ll \sigma G^2$ contain the partial contribution of initial modes with $m = +|m|$ and $m = -|m|$ of the equal amplitude. Therefore, they have the structure of *angular standing waves*, and the magnetization for them is proportional to $\cos(|m|\chi) \cdot \exp(i\omega t)$ and $\sin(|m|\chi) \cdot \exp(i\omega t)$, instead of $\exp(i\omega t \pm i |m| \chi)$ for modes with definite $m$.

It is worth noting that for both limit cases, described by Eqs. (37, 39), that the splitting of collective modes originating from the doublet with $m = +1$ and $m = -1$ for the isolated dot appears. This splitting is quadratic over the interaction for $\omega_0 - \bar{\omega}_0 > \sigma G^2$, and even higher (linear in the interaction), for small initial splitting $\omega_0 - \bar{\omega}_0 \ll \sigma G^2$. Thus, the contribution to the doublet splitting caused by the last item in (37) can not be neglected. For the single dot such splitting emerges only due to the gyroscopic character of the vortex motion and is small, being proportional to the ratio $\Delta_0/R$. Here the splitting is connected to the lattice effects only. In other words, the symmetry of the square lattice does not support exact solutions with well-defined azimuthal number *m*. As a first result of this symmetry breaking, the standing waves with defined value of |*m*| appear as normal modes. The common effect has been seen in numerical simulations of the normal modes for lattice systems with high enough easy-plane anisotropy [50]. It is also noted that from these numerical results the effects of lattice splitting are strongest for the modes corresponding to the square symmetry of the lattice, such as modes with |*m*| = 2, 4 and so on. The detailed analysis of doublets with |*m*| > 1, not observed in experiment yet, and is far from the aim of this article. We will only mention that for such modes the formation of angular standing waves with angular dependence of form cos(|*m*|χ) and sin(|*m*|χ)·, with |*m*| crossing node lines for each mode, is expected.

One more problem that can be analyzed on the basis of the Hamiltonian (33), is related to the influence of higher modes on low-frequency branches of excitations through dipole coupling. Specifically, assuming that the precessional vortex mode corresponds to the mode with *m* = 1, and one of the higher frequency modes, $\bar{\omega}_0 \gg \omega_0 \equiv \omega_{PM}$ corresponds to *m* = – 1, then one can estimate qualitatively how the higher frequency mode will affect the spectrum of excitations corresponding to collective oscillations of vortex cores. For this case the inequality $\bar{\omega}_0 \gg \omega_0 = \omega_{PM}$ gives the spectrum of the collectivized vortex precessional mode $\omega_{PM}(\vec{k})$ in the following form

$$\omega_{PM}(\vec{k}) = \omega^{(0)}_{PM}(\vec{k})[1 - G^2 \bar{G}^2(\sigma^2 - |3\sigma_c|^2)/2\bar{\omega}_0^2] ,  \qquad (40)$$

where $\omega^{(0)}_{PM}(\vec{k})$ determines the spectrum of vortex precessional mode without taking into account interaction with the higher modes given by Eq. (20). One can see that the corresponding correction is small, the order of the parameter ($\omega_0$, $G^2\sigma$)/ $\bar{\omega}_0$, and in order to describe the translation mode Eq. (22) can be used.

### 6. Summary and concluding remarks

In conclusion, the dynamic properties of the array of vortex state magnetic dots without direct exchange interaction between dots have been investigated. We limit ourselves to the simplest case with all the array dots in the same state having equal vorticity and polarization. It is useful to discuss the possibility of realization of such state. It is possible to prepare a dot array with the parallel vortex polarizations by use of the magnetic field perpendicular to the plane. The same vorticity for each dot can be created by electric current flowing perpendicularly to the dot plane, as it was proposed in Ref. [51]. For such systems, only the magnetic dipole interaction can be a source of dot interaction. In fact, it has to be considered as a nonlocal interaction having two types of magnetization inhomogenities; the first originating in the patterned array structure, and the second arising from possible non-uniform distribution of magnetization within a single dot. Thus, the magnetic dipole interaction is taken as a sole source of the interaction between individual dots, but the approach proposed here goes beyond the dipolar approximation, within which the single dot is considered as point magnetic dipole.

We have shown that there are new and unexpected effects that appear in the collective mode spectrum. There are significant differences depending mostly on properties of the single dot mode. The results differ strongly for dipolar active modes, corresponding to the oscillations of total magnetic moment of a dot, and non-dipolar modes, for which the oscillations of magnetization are such that the total magnetic moment remains zero. For the dipolar modes, the dispersion relation is non-analytic as $\vec{k} \to 0$ because of the long-range nature of the dipolar interaction of oscillating magnetic moments of dots, and the presence of singularities in dipolar sums. For such modes the magnon spectra have the peculiarity of the type $\omega(\vec{k}) - \omega_g \propto |\vec{k}|$ as $\vec{k} \to 0$, where $\omega_g$ is a nonzero gap frequency. For non-dipolar modes the singularities are absent and long-wave asymptotics become standard, $\omega(\vec{k}) - \omega_g \propto |\vec{k}|^2$. Obviously, when taking into account a non-dipolar interaction in addition to the dipolar interaction, its role is negligible in the long wave limit, but the non-dipolar contribution can be significant for non-small $|\vec{k}|a \approx 1$, and even smaller values if the dipolar interaction is small for some reason. An example of such behavior is the collective mode originating from radially-symmetric $m = 0$ modes for vortex state dots (see Section 4).

An important and non-trivial property of collective $m = 0$ modes for vortex state dots is the decreasing dependence of the mode frequency, $\omega(\vec{k})$ on the wave vector $\vec{k}$. An interesting potential application can result from this mode because of the nature of this dispersion relation. Its group velocity, $\partial\omega/\partial k$ is negative for all values of $k$ similar to backward volume magnetostatic waves that can propagate in magnetic thin films. Therefore, it might be possible to use a pulse perpendicular to the array to generate spectrally wide envelope pulses in these systems that can be observed using imaging techniques. Observation of the time development of the envelope will give information about the dispersion, $\partial^2\omega/\partial k^2$, diffraction effects arising from the direction perpendicular to propagation, and nonlinearities.

Here we obtained full spectra for the change of quasimomentum within all of the Brillouin zone. The direct measurement of the dependence $\omega(\vec{k})$ can be done by the Brillouin light scattering method [3]. Previously, only spectra with small dispersion have been described in the literature. Our calculations do demonstrate that the dispersion for typical arrays of thin dots, even for small separation, is small, but we have remarked that the dispersion is a strongly increasing function of the dot thickness. It will be interesting to observe non-monotonic dependence $\omega(\vec{k})$ for $m = 0$ collective modes.

The concrete calculations here have been done for the most symmetric configurations of the system; the ideal square lattice array, and individual dots having rotational uniaxial symmetry. For the vortex state dots, an interesting symmetrical effect appears in the dot interaction. The symmetry of an array lattice is lower than for a single dot of circular shape. Because of this, the symmetry of modes will also be lowered as a result of the interaction. For the single vortex state dot almost degenerate doublets with azimuthal numbers $m = \pm |m|$ are known to be present. The lowering of symmetry caused by dot interaction in the lattice produces angular standing waves with the magnetization proportional to $\cos(|m|\chi)\cdot\exp(i\omega t)$ and $\sin(|m|\chi)\cdot\exp(i\omega t)$, instead of $\exp(i\omega t \pm i|m|\chi)$ for modes with definite $m$ values forming doublets for the single vortex state dot.

**Appendix.**

The expressions for collective modes frequencies contain series such as $\sum_{|\vec{l}|\neq 0} e^{i\vec{k}\vec{l}}/|\vec{l}|^{3+2p}$, where $p = 0$ for the dipole sums $\sigma(\vec{k})$ and $\sigma_c(\vec{k})$, which is important for analysis of the $m = 0$ and $m = \pm 1$ modes, and $p = 1, 2, ...$ for the rest of the modes. Here and below in this Section we will use the dimensionless vector $\vec{l}$ and the condition $\vec{l} \neq 0$ in the sums is implied.

Let us discuss properties of these series. As we will demonstrate these series for $p = 0$ yield the properties fundamentally different from those for series with $p > 0$. The point is that the double sum converges rather slowly. This manifests especially in sum properties near the origin, $\vec{k} = 0$. Analyzing small deviations from these points, considering small $\vec{k}$, one has to calculate derivatives like $[\partial^2\sigma(\vec{k})/\partial k_i \partial k_j]$ at the point $\vec{k} = 0$. Term by term differentiation of the dipole sums gives series, which are alternating and converge only conditionally. For example, for $\vec{k} = 0$, i.e., in the physically most interesting case of long wave oscillations, the corresponding coefficient of $\vec{k}^2$ is described by the divergent series $\sum 1/|\vec{l}|$. Therefore, the dipole sums can be non-analytical and it results in a non-standard dispersion law for collective oscillations.

In the case $p > 0$ the situation is more conventional: as $|\vec{k}| \to 0$ the corresponding series

$$\sum \frac{e^{i\vec{k}\vec{l}}}{|\vec{l}|^{3+2p}} - \sum \frac{1}{|\vec{l}|^{3+2p}} = -q^2 D = -q^2 \sum \frac{1}{|\vec{l}|^{1+2p}} \qquad (41)$$

are convergent and collective oscillations with $p > 0$ are characterized as $|\vec{k}| \to 0$ by a standard quadratic dispersion law. In this case the peculiarities in higher derivatives emerge, which may also produce non-analytical expressions like $\sigma_2(\vec{k}) = D\vec{k}^2 + D'|\vec{k}|^3$; however, observation of such effects is questionable.

Next consider the sum, $\sigma_{2p}(\vec{k}) = \sum e^{i\vec{k}\vec{l}}/|\vec{l}|^{3+2p}$. The numerical data for the dependence of $\sigma_{2p}(\vec{k})$ at $p = 0, 1, 2$ and $3$ inside the first Brillouin zone for two symmetrical directions of the quasimomentum, $\vec{k}$ are presented in Fig. 5. The principal difference of the $\vec{k}$ dependence of these sums around the origin is clearly seen. For the sum $\sigma(\vec{k})$ with $p = 0$ (solid curve on the figure) the sharp maximum with non-analytic dependence is present at $\vec{k} = 0$, whereas for other series with $p \neq 0$, represented by symbols in the figure, the maximum looks like the usual parabolic dependence. The behavior of all the functions looks much more similar for large enough values of $|\vec{k}|/k_B$. At non-zero $p$ all the functions $\sigma_{2p}(\vec{k})$ are similar for all $|\vec{k}|$, including the domain near the origin, and at asymptotically large $p$ all of them are described by the nearest neighbor approximation, $\sigma_{2p}(\vec{k}) \to 2[\cos(k_x a) + \cos(k_y a)]$ as $p \to \infty$. The numerical analysis shows that this approximation gives the accuracy better than 10% for all $p > 1$. For $p = 1$ the next - nearest neighbor terms are of the order 20 %. Including these terms, one can present this series as

$$\sigma_2(k) = 2[\cos(k_x a) + \cos(k_y a)] + (1/\sqrt{2})\cos(k_x a)\cos(k_y a) , \qquad (42)$$

with the accuracy better than 8 % .

Thus, the principal difference between $\sigma(\vec{k})$ and other series $\sigma_{2p}(\vec{k})$ at $p \neq 0$ is caused by slow convergence of $\sigma(\vec{k})$. The same property is present for the complex sum of the type $\sigma_c(\vec{k})$, which is also important for the description of dipole-coupled modes.

Acknowledgement

This research was supported by the National Science Foundation under Grants Number DMR-9974273 and DMR-9972507.

**Figure Captions**

Fig. 1. The structure of the dot array, and definition of coordinate systems. The gray circles represent ferromagnetic elements (magnetic dots), which are considered as thin cylinders.

Fig. 2. The dependence of the dipole sums $\sigma(0) - \sigma(\vec{k})$ and $\sigma_c(\vec{k})$ on the quasimomentum $\vec{k}$ for symmetric directions of the square lattice within the first Brillouin zone. Here and below $k_B$ is the maximal value of the wave vector modulus for a given direction, corresponding to the boundary of the Brillouin zone, $k_B = \pi/a$ for $\vec{k} \parallel (1,0)$ and $k_B = \sqrt{2}\,\pi/a$ for $\vec{k} \parallel (1,1)$.

Fig. 3. The collective mode frequencies for the vortex precessional mode (in the units of the frequency for isolated dot $\omega_{PM}$) for the vortex-state dot arrays of different densities. The solid curve represents the data for a dense array with $a = 1.1\cdot(2R)$, the dotted curve corresponds to $a = 1.5\cdot(2R)$, and the dashed curve gives the data for a low density array with $a = 4R$.

Fig. 4. The frequencies of the lowest collective mode with $m = 0$ (in the units of the frequency for isolated dot, $\omega_0$) for the dense array with $a = 1.1(2R)$ of the vortex-state dots for two different values of the aspect ratio $L/R$.. For small aspect ratio, $L/R = 0.2$, the region near the origin $\vec{k} = 0$ is shown in the left top insert. Here the data for the high value aspect ratio, $L/R = 1$, (for such high *L/R* our equations are valid for qualitative estimate only) is present for the demonstration that even at extremely high *L/R* the contribution of non-dipolar interaction is important.

Fig. 5. The dependence $\sigma_{2p}(\vec{k})$ for the sums with different values of *p*. The dipole sum $\sigma(\vec{k})$ with *p* = 0 is represented by the solid curve, the sum $\sigma_2(\vec{k})$ with *p* = 1, appearing for non-dipole interaction, see Sec. 4, is represented by solid line with circles. The higher sums with *p* = 2, 3 are depicted by up and down triangles, respectively. The dashed curve is the result of nearest neighbor approximation.

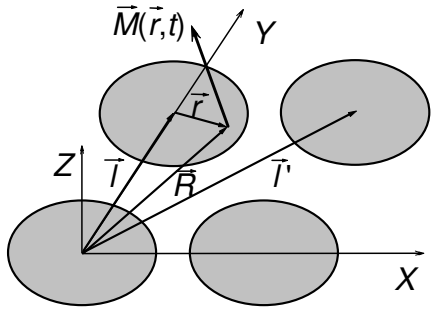

Fig. 1

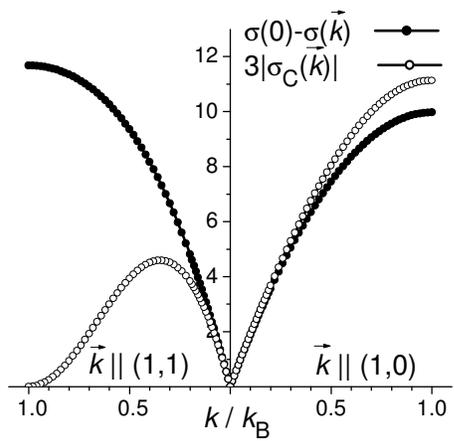

Fig. 2

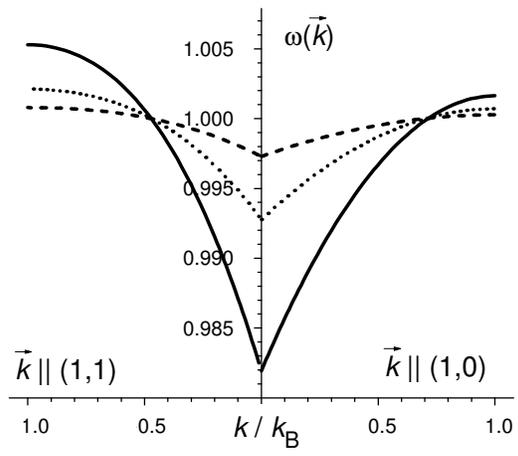

Fig. 3

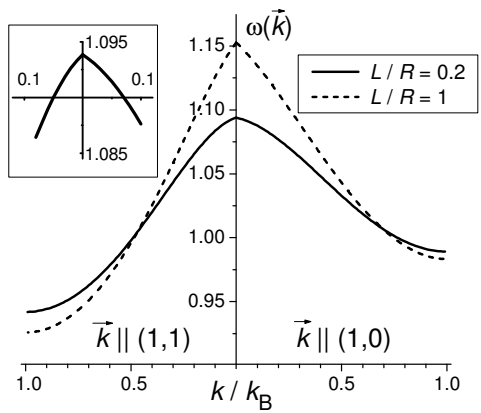

Fig. 4

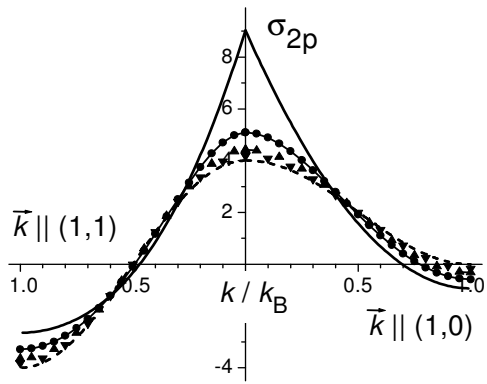

Fig. 5